\documentclass[aps,pra,twocolumn,superscriptaddress,amssymb,showpacs]{revtex4}

\usepackage{graphicx}

\begin{document}


\title{Entangled light in transition through the generation threshold}


\author{G.~Yu.~Kryuchkyan}
\email[]{gkryuchk@server.physdep.r.am}
\affiliation{Yerevan State University, A. Manookyan 1, 375049,
Yerevan, Armenia} \affiliation{Institute for Physical Research,
National Academy of Sciences,\\Ashtarak-2, 378410, Armenia}

\author{L.~A.~Manukyan}
\email[]{leman@ipr.sci.am}
\affiliation{Institute for Physical Research, National Academy of
Sciences,\\Ashtarak-2, 378410, Armenia}


\begin{abstract}
We investigate continuous variable entangling resources on the
base of two-mode squeezing for all operational regimes of a
nondegenerate optical parametric oscillator with allowance for
quantum noise of arbitrary level. The results for the quadrature
variances of a pair of generated modes are obtained by using the
exact steady-state solution of Fokker-Planck equation for the
complex P-quasiprobability distribution function. We find a simple
expression for the squeezed variances in the near-threshold range
and conclude that the maximal two-mode squeezing reaches 50\%
relative to the level of vacuum fluctuations and is achieved at
the pump field intensity close to the generation threshold. The
distinction between the degree of two-mode squeezing for
monostable and bistable operational regimes is cleared up.
\end{abstract}


\maketitle

Recent years have witnessed an increasing interest in quantum
information theory dealing with entanglement of continuous
variable (CV) systems \cite{QITwithCV}. CV entangled states of
light hold the key for quantum communications at the
high-intensity level, as CV analogues of various protocols
developed originally in the framework of discrete quantum
variables have been established (see, for example
\cite{CVexperimental,Ralph}). An important example provides
entangled Einstein-Podolski-Rosen (EPR) state of orthogonally
polarized but frequency degenerate beams, efficiently generated
via nonlinear optical process of parametric down-conversion. It
has been pointed out in \cite{Reid} and has been demonstrated
experimentally in \cite{OuPereiraKimblePeng} for CV, employing
sub-threshold nondegenarate optical parametric oscillator (NOPO).
Then a CV entanglement source was built from two independent
quadrature squeezed beams combined on a beam-splitter
\cite{CVexperimental}. The generation of CV EPR entanglement using
an optical fibre interferometer is also demonstrated
\cite{Korolkova}. It should be noted that so far most experimental
realizations of the CV\ entanglement on NOPO's have only been
operated below threshold. A natural next step is the extension of
these investigations to laser-like systems generating entangled
bright-light states. Unfortunately, both theoretical and
experimental studies of these problems are very complicated and
only rare examples are known up to now. Many efforts have been
devoted to the study of intensity correlated twin beams from NOPO
above threshold \cite{Mert}. The conditional generation of
sub-Poissonian light from bright twin beams in the CV regime is
experimentally demonstrated in \cite{Lauret}. Further experimental
studies of bright two-mode entangled state from cw nondegenerate
optical parametric amplifier have been made in Refs.\
\cite{Zhang}. The generation of CV polarization entanglement by
mixing a pair of polarization squeezed beams is realized in
\cite{PRA_65_052306}.

The purpose of this paper is to investigate physical properties
and the presence of CV\ entanglement for NOPO in its transition
through the generation threshold as well as in the regime of
lasing. One of the principal problems in this study is the
description of quantum fluctuations. In most theoretical works
nonclassical effects and entanglement resources of nonlinear
quantum systems are usually described within linear treatment of
quantum noise. It is obvious that such approach does not describe
the critical ranges (threshold, point of multistability, etc)
where the level of quantum noise increases substantially. We use a
more adequate approach within the framework of exact nonlinear
treatment of quantum fluctuations via the solution of the
Fokker-Planck equation for the quasiprobability distribution
function. Deriving the quasiprobability functions so far has been
performed only for a few simple models (see \cite {KK} and
references therein). For the NOPO the exact steady-state solution
of the Fokker-Planck equation in the complex $P$-representation
first was obtained in \cite{McN}. We will use the generalized form
of such solution \cite{KP} which involves also detunings of the
modes and hence admits bistability.

There are various questions that emerge in the study of these
problems. Because NOPO displays both monostable and bistable
regimes it is important to understand how the properties of
entanglement depend on the operational regimes. What are
peculiarities of entanglement in the vicinity of threshold? Will
entanglement take place in the regime of lasing or how far it can
be extended into the high intensity domain? Answering these
questions is extremely important for a deeper understanding of
quantum entanglement, and also from the perspective of creating an
entangled light laser.

We consider a type-II phase matched NOPO with triply resonant
cavity that supports the pump mode at the frequency $\omega _{3}$
and two orthogonally
polarized modes at the same frequency $\omega _{1}=\omega _{2}=\omega _{3}/2$%
. The pump mode is driven by coherent field at the frequency
$\omega _{L}\simeq \omega _{3}$. The relevant interaction
Hamiltonian is
\begin{eqnarray}
\lefteqn{H=i\hbar \kappa \left( a_{1}^{+} a_{2}^{+} a_{3} - a_{1}
a_{2} a_{3}^{+} \right) + \mbox{} } \label{Hamilt}\\* %
&&\mbox{}+i\hbar \left( Ee^{-i\omega _{L}t}a_{3}^{+}-E^{\ast
}e^{i\omega_{L}t}a_{3}\right)+
\sum_{i=1}^{3}\left(a_{i}\Gamma_{i}^{+}+a_{i}^{+}\Gamma_{i}\right)
\nonumber
\end{eqnarray}
where $a_{i}$ ($i=1,2,3$) are the operators of the modes $\omega _{i}$, $%
\kappa $ is the coupling constant, $E=\left| E\right| e^{i\Phi
_{E}}$ is the complex amplitude of the pump field. The last term
in (\ref{Hamilt}) describes mode damping in the cavity in terms of
the reservoir operators $\Gamma_{i}$ and $\Gamma_{i}^{+}$, which
determine the damping factors $\gamma_{i}$. We take into account
the detunings of the cavity $\Delta _{3}=\omega _{L}-\omega _{3}$,
and $\Delta _{1}=\Delta _{2}=\Delta =\omega _{L}/2-\omega _{1,2}$.
We also assume perfect symmetry between the orthogonally polarized
modes provided that they decay at the same rate:\ $\gamma
_{1}=\gamma _{2}=\gamma $.

It is well known that linear treatment of NOPO is not
self-consistent due to the phase diffusion. According to this
effect the difference between the phases of the signal and idler
modes, as well as each of the phases can not be defined in the
above threshold regime of generation. On the whole, the well
founded linearization procedure cannot be applied for this system.
Nevertheless, the linearization procedure and analysis of quantum
fluctuations for NOPO become possible due to the additional
assumptions about temporal behavior of the difference between the
phases of the generated modes \cite{Drummond}. Within this
assumption the two-mode squeezing spectra are obtained for both
below- and above-threshold regimes. These results are inapplicable
in the near-threshold as well as in the bistable ranges of
generation, where a more accurate treatment of quantum noise is
necessary. Our analysis of NOPO is based on the steady-state exact
solution of the Fokker-Planck equation. It is obvious, that in the
framework of this approach we avoid both difficulties of the
linear theory. The price one has to pay for this advantage is the
impossibility to perform temporal description of quantum
fluctuations and hence squeezing spectra cannot be calculated.
Nevertheless, our analysis does not require spectral description
of squeezing and therefore we do not encounter any principal
difficulties.

We start with the Fokker-Planck equation of the system in the so
called complex P-representation of the density matrix and in the
case of high cavity losses for the pump mode ($\gamma _{3}\gg
\gamma $), however, in the operational regime the pump depletion
effects are involved:
\begin{widetext}
\begin{eqnarray}
\frac{\partial}{\partial t}P(\overline{\alpha}, t)&=& \left\{ 
-\frac{\partial}{\partial\alpha_{1}}\left[
-\overline{\gamma}_{1}\alpha_{1}+\kappa
\left(\frac{E-\kappa\alpha_{1}\alpha_{2}}{\overline{\gamma}_{3}}\right)
\alpha_{2}^{+}\right] 
-\frac{\partial}{\partial\alpha_{1}^{+}}\left[
-\overline{\gamma}_{1}^{\ast}\alpha_{1}^{+}+\kappa
\left(\frac{E^{\ast}-\kappa\alpha_{1}^{+}\alpha_{2}^{+}}{\overline{\gamma}_{3}^{\ast}}\right)
\alpha_{2}\right] \right.
 \nonumber \\
&&\mbox{}-\left.\frac{\partial}{\partial\alpha_{2}}\left[
-\overline{\gamma}_{2}\alpha_{2}+\kappa
\left(\frac{E-\kappa\alpha_{1}\alpha_{2}}{\overline{\gamma}_{3}}\right)
\alpha_{1}^{+}\right] 
-\frac{\partial}{\partial\alpha_{2}^{+}}\left[
-\overline{\gamma}_{2}^{\ast}\alpha_{2}^{+}+\kappa
\left(\frac{E^{\ast}-\kappa\alpha_{1}^{+}\alpha_{2}^{+}}{\overline{\gamma}_{3}^{\ast}}\right)
\alpha_{1}\right] \right.
\nonumber \\
&&\mbox{}+\left. \kappa
\frac{\partial^{2}}{\partial\alpha_{1}\partial\alpha_{2}}
\left[\frac{E-\kappa\alpha_{1}\alpha_{2}}{\overline{\gamma}_{3}}\right]
+ \kappa
\frac{\partial^{2}}{\partial\alpha_{1}^{+}\partial\alpha_{2}^{+}}
\left[\frac{E^{\ast}-\kappa\alpha_{1}^{+}\alpha_{2}^{+}}{\overline{\gamma}_{3}^{\ast}}\right]
\right\} P(\overline{\alpha}, t).  \label{Fokker-Planck} %
\end{eqnarray}
\end{widetext}
Here $\overline{\alpha}=(\alpha_{1}, \alpha_{1}^{+}, \alpha_{2},
\alpha_{2}^{+})$, $\alpha_{i}$, $\alpha_{i}^{+}$ ($i=1,2$) are the
independent complex variables corresponding to the operators
$a_{i}$, $a_{i}^{+}$, and
$\overline{\gamma}_{j}=\gamma_{j}-i\Delta_{j}$,
($\overline{\gamma}_1=\overline{\gamma}_2=\overline{\gamma}$).

The normally-ordered moments of time-dependent operators are
calculated through the P-quasiprobability distribution function as
\begin{eqnarray}
\lefteqn{\left\langle a_{1}^{+}(t)^{k} a_{1}(t)^{l}
a_{2}^{+}(t)^{m} a_{2}(t)^{n} \right\rangle = } \\*%
&&= \int d \alpha_{1}^{+} d \alpha_{1} d \alpha_{2}^{+} d
\alpha_{2} P(\overline{\alpha}, t)
\alpha_{1}^{+k}\alpha_{1}^{l}\alpha_{2}^{+m}\alpha_{2}^{n}.
\nonumber
\end{eqnarray}

Below we consider only the stationary steady-state regime and drop
the time dependence of operators when calculating one-time
expectation values. Using the steady-state solution of the
equation (\ref{Fokker-Planck}), obtained in \cite{KP}, and the
method of integration in the complex plane \cite{McN} we find:
\begin{widetext}
\begin{eqnarray}
&&\left\langle a_{1}^{+k}a_{1}^{l}a_{2}^{+k}a_{2}^{l}\right\rangle =\frac{%
\varepsilon ^{l}\varepsilon ^{\ast k}}{N}\frac{l!k!}{(\Lambda
+1)_{l}(\Lambda ^{\ast }+1)_{k}} \sum_{j=0}^{\infty }\frac{\left(
l+1\right) _{j}\left( k+1\right) _{j}}{\left( \Lambda +l+1\right)
_{j}\left( \Lambda ^{\ast }+k+1\right)_{j}}\frac{p^{j}}{(j!)^{2}},
\label{momentklkl}\\ %
&&\left\langle a_{1}^{+m}a_{1}^{m}a_{2}^{+n}a_{2}^{n}\right\rangle
= \frac{p^{m}}{2^{m+n}N}\left| \frac{m!}{(\Lambda +1)_{m}}\right|
^{2} \sum_{j=0}^{\infty }\left| \frac{\left( m+1\right)
_{j}}{\left( \Lambda +m+1\right) _{j}}\right|
^{2}\frac{p^{j}}{j!(j+m-n)!}, \text{if }m\geq n,
\label{momentmmnn} \\ %
&&\left\langle a_{1}^{+m}a_{1}^{m}a_{2}^{+n}a_{2}^{n}\right\rangle
=\left\langle a_{1}^{+n}a_{1}^{n}a_{2}^{+m}a_{2}^{m}\right\rangle
, \text{if } n>m,
\end{eqnarray}
\end{widetext}
where $\varepsilon=E/\kappa$, $p=\left| 2\varepsilon \right| ^{2}$
is the scaled pump intensity, $\Lambda =2\overline{\gamma
\gamma}_{3}/\kappa ^{2}$, $(x)_{j}:=x(x+1)...(x+j-1)$,
$(x)_{0}=1$, and
\begin{equation}
N=\sum_{j=0}^{\infty }\frac{p^{j}}{\left| \left( \Lambda +1\right)
_{j}\right| ^{2}}=\sum_{j=0}^{\infty }N_{j}. \label{Nexpression}
\end{equation}
In addition to these expressions we note that the system under
consideration has the following property, conditioned by its most
general symmetries:
\begin{equation}
\left\langle
a_{1}^{+k}a_{1}^{l}a_{2}^{+m}a_{2}^{n}a_{3}^{+p}a_{3}^{q}\right\rangle
=0, \text{\ if\ }k-l\neq m-n.  \label{moment_klmnpq}
\end{equation}

This property is the consequence of the rotational symmetry of the
Hamiltonian (\ref{Hamilt}) $U\left( \theta \right) HU^{-1}\left(
\theta \right) =H$ for any $\theta $, where $U\left( \theta
\right) =\exp \left[ i\theta \left(
a_{1}^{+}a_{1}-a_{2}^{+}a_{2}\right) \right] $. Since the Lindblad
part of the master equation is invariant with respect to such
transformation too, then the steady state density operator of NOPO
must commute with $U\left( \theta \right) $: $U\left( \theta
\right) \rho U^{-1}\left( \theta \right) =\rho $, whereupon we
arrive at the relation (\ref{moment_klmnpq}).

We will use only the photon number and the $\left\langle
a_{1}a_{2}\right\rangle $ moment in the further discussion, so we
write their expressions explicitly:
\begin{eqnarray}
&&n=\left\langle a_{1}^{+}a_{1}\right\rangle =\left\langle
a_{2}^{+}a_{2}\right\rangle =\frac{1}{2N}\sum_{j=1}^{\infty }\frac{jp^{j}}{%
\left| \left( \Lambda +1\right) _{j}\right| ^{2}},
\label{meanphotonexact}\\
&&\left\langle a_{1}a_{2}\right\rangle =\frac{e^{i\Phi _{E}}}{2N\sqrt{p}}%
\sum_{j=1}^{\infty }\frac{j\left( \Lambda ^{\ast }+j\right)
p^{j}}{\left| \left( \Lambda +1\right) _{j}\right| ^{2}}.
\label{a1a2}
\end{eqnarray}

We note that the state generated in NOPO is non-Gaussian one, i.e.
its Wigner function is non-Gaussian \cite{NOPOwigner}. So far, the
inseparability problem for bipartite non-Gaussian states is far
from being understood. On the theoretical side, the necessary and
sufficient conditions for the separability of bipartite CV systems
have been fully developed only for Gaussian states \cite{Duan}. To
characterize CV entanglement we have chosen the inseparability
criterion based on the total variance of a pair of EPR type
operators. For a pair of optical beams generated in NOPO this
criterion characterizes the entanglement in terms of quadrature
operators
\begin{eqnarray}
X_{k}&=&X_{k}(\theta_{k}, t)=\frac{1}{\sqrt{2}}\left[
a_{k}^{+}(t)e^{
-i\theta _{k}}+a_{k}(t)e^{i\theta _{k}}\right], \nonumber \\
Y_{k}&=&Y_{k}(\theta_{k}, t)=X_{k}(\theta_{k}-\frac{\pi}{2}, t),
\label{quadrature_amplitudes}
\end{eqnarray}
($k=1,2$) and due to the mentioned symmetries is reduced to the
following form:
\begin{equation}
V(\theta_{1}, \theta_{2}):=V\left( X_{1}-X_{2}\right) \equiv
V\left( Y_{1}+Y_{2}\right) <1\text{.} \label{r_criterion}
\end{equation}

Here $V(x)$ is a convenient short-hand notation for the variance $%
\left\langle x^{2}\right\rangle -\left\langle x\right\rangle
^{2}$. Inequalities (\ref{r_criterion}) require the variances of
both conjugate variables $X_{1}-X_{2}$ and $Y_{1}+Y_{2}$ to drop
below the level of vacuum fluctuations. Since the states of NOPO
are non-Gaussian, the criterion (\ref{r_criterion}) is only
sufficient for inseparability.

At this point we must note the difference between the focus of our
paper and most of the preceding work devoted to the study of
two-mode squeezing. It is an established standard to describe
squeezing with the spectra of quantum fluctuations of the
considered variables. The recent experiment on spectral
investigation of criteria for CV entanglement was presented in the
paper \cite{Bowen_PRL_90}. Unlike that, we aim to analyze the
separability properties of the system solely with the help of the
criterion (\ref{r_criterion}), which involves a single integral
characteristic of the NOPO quantum state. For the completeness of
the results it is useful to remind how intracavity variances are
related to the spectra of squeezing of output fields. For this,
let us consider the special scheme of generation, when the
couplings of in- and out-fields occur at only one of the
ring-cavity mirrors. For the case when only the fundamental mode
is coherently driven by the pump field, while subharmonic modes
are initially in the vacuum state, we have for the output fields
$a_{i}^{out}(t)=\sqrt{2\gamma }a_{i}(t), (i=1,2)$. The spectra of
two-mode quadrature amplitude squeezing is
\begin{eqnarray}
\lefteqn{S(\omega , \theta_{1}, \theta_{2})=} \\*
&&=\int_{-\infty}^{\infty}e^{i\omega \tau} \langle X_{-}^{out}(
\theta_{1}, \theta_{2}, t)X_{-}^{out}(\theta_{1}, \theta_{2},
t+\tau)\rangle d\tau . \nonumber
\end{eqnarray}

Here $X_{-}^{out}(\theta_{1}, \theta_{2}, t) = \sqrt{2\gamma
}\left[ X_{1}(\theta_{1}, t) - X_{2}(\theta_{2}, t) \right] $, and
$t$ is sufficiently large ($t\gg\gamma^{-1}$) to ensure steady
state, so that $S(\omega , \theta_{1}, \theta_{2})$ does not
depend on $t$ .

The variance $V(\theta_{1}, \theta_{2})$ is expressed with the
above spectra in the following way:
\begin{equation}
2\gamma V(\theta_{1}, \theta_{2}) = \frac{1}{2\pi }
\int_{-\infty}^{\infty} S(\omega , \theta_{1}, \theta_{2})
d\omega.
\end{equation}

Let us turn to the integral variances. With an appropriate
selection of the phases $\theta _{k}$ of the quadrature
amplitudes, namely
\begin{equation}
\theta _{1}+\theta _{2}=\arg \left\langle a_{1}a_{2}\right\rangle,
\label{minimizing_phases}
\end{equation}
we arrive at the following minimum value of $V(\theta_{1},
\theta_{2})$
\begin{equation}
V_{\min }=1+2\left( n-\left| \left\langle a_{1}a_{2}\right\rangle
\right| \right) \text{.}  \label{r_expression}
\end{equation}

Detailed physical analysis of the variance (\ref{r_expression}) on
the base of the exact expressions (\ref{meanphotonexact}),
(\ref{a1a2}) seems to be a rather complicated task. To proceed
further we transform the formula (\ref{a1a2}) to a more
appropriate form
\begin{equation}
\left\langle a_{1}a_{2}\right\rangle =e^{i\Phi _{E}}
\frac{n}{\sqrt{p}}\left( \Lambda ^{\ast
}+2n+\frac{p}{n}\frac{\partial n}{\partial p}\right) .
\label{a1a2_altform}
\end{equation}

What is important is that $\left\langle a_{1}a_{2}\right\rangle $
is expressed with $n$ in a simple enough form, much easier to
understand than its expansion in powers of $p$. The advantage of
that form is due to the intuitiveness of the behavior of $n$ in
various feasible operational regimes. Moreover, semiclassical
solution for the mean photon number, where applicable, can be
employed to obtain a semiclassical solution for $\left\langle
a_{1}a_{2}\right\rangle $. To this end, we remind relevant results
of the semiclassical approximation. In the regime below threshold
$p<p^{th}$ the excitations of modes are at the level of
spontaneous noise, and above threshold they have the form
\begin{equation}
n_{cl}=\frac{1}{2}\left( -%
\mathop{\rm Re}%
\Lambda +\sqrt{p-\left(
\mathop{\rm Im}%
\Lambda \right) ^{2}}\right) \text{.}  \label{meanphotonclassical}
\end{equation}

\begin{figure}[tbp]
\includegraphics[angle=-90,width=0.48\textwidth]{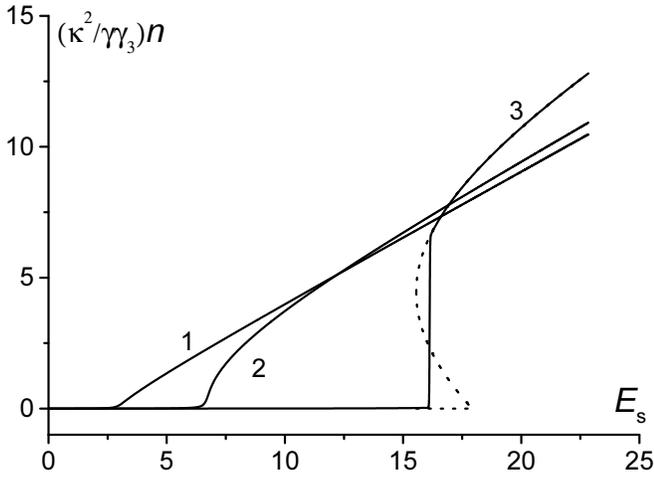}
\caption{Scaled mean photon number versus dimensionless amplitude
of the pump field $E_{s}=(2\protect\kappa /\protect\gamma
\protect\gamma _{3})\left| E\right| $ for monostable (curve 1,
$\Delta /\protect\gamma =1$), interjacent (curve 2, $\Delta
/\protect\gamma =3$) and bistable (curve 3, $\Delta
/\protect\gamma
=7$) regimes ($\protect\kappa /\protect\gamma =0.5$, $\protect\gamma _{3}/%
\protect\gamma =18)$. Dot curve visualizes the classical analysis
for the bistable case.} \label{intensity_fig}
\end{figure}

\begin{figure}
\includegraphics[angle=-90,width=0.48\textwidth]{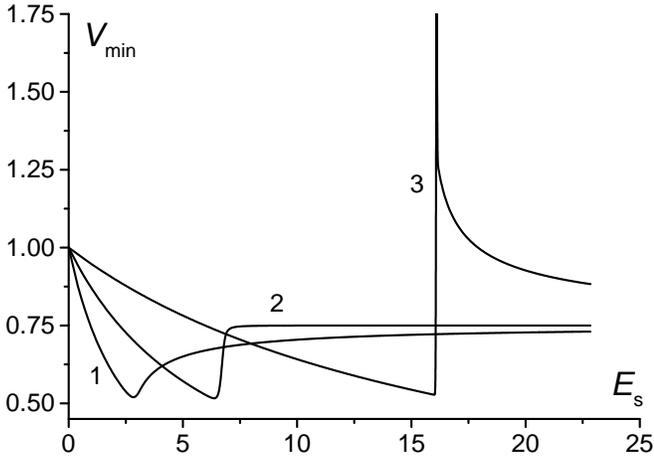}
\caption{Minimized variance $V\left( X_{1}-X_{2}\right) $ versus
dimensionless amplitude of the pump field $E_{s}$ for the same
parameters as for Fig.\ \ref{intensity_fig}.} \label{variance_fig}
\end{figure}

The case $\mathop{\rm Re}\Lambda >0$ corresponds to monostable dynamics with threshold $%
p_{m}^{th}=\left| \Lambda \right| ^{2}$. (The corresponding
threshold value of the pump field $E^{th}=\left| \left( \gamma
-i\Delta \right) \left(
\gamma _{3}-i\Delta _{3}\right) \right| /\kappa $.) The opposite case $%
\mathop{\rm Re}%
\Lambda <0$ is bistable with threshold $p_{b}^{th}=\left(
\mathop{\rm Im}%
\Lambda \right) ^{2}$ and the stability region of the zeroth
solution extending up to $p=\left| \Lambda \right| ^{2}$. For
completeness, we present in Fig.\ \ref{intensity_fig}
quantum-mechanical results for mean photon number of the modes
(\ref{meanphotonexact}) for three values of
detunings corresponding to monostable, bistable and interjacent ($%
\mathop{\rm Re}%
\Lambda =0$) regimes. For the bistable dynamics the critical
region, i.e. the range of the pump intensity, where the system
passes onto the nontrivial
classical branch, lies between $p_{b}^{th}$ and $\left| \Lambda \right| ^{2}$%
. The quantum critical region of monostable dynamics is in the
vicinity of the generation threshold $p_{m}^{th}=\left| \Lambda
\right| ^{2}$.

We are now in a position to study the entanglement effects and
will state the main results of the paper. What is important is
that we find the quantities (\ref{meanphotonexact})-(\ref{a1a2})
through the exact steady-state solution of the Fokker-Planck
equation and carry out an exact quantum statistical analysis.

Fig.\ \ref{variance_fig} illustrates the dependence of $V_{\min }$
on the pump amplitude. One can see that for monostable dynamics
(curve 1) entanglement is realized in the entire range of pump
intensities. In all cases maximal degree of two-mode squeezing
$V_{\min }=0.5$ is achieved within the critical region. Above the
critical point mean photon numbers of the modes increase
considerably and variance $V_{\min }$ starts to increase too. For
the bistable case (curve 3) growth of $V_{\min }$ is much faster
and larger so that the sufficient criterion for inseparability
(\ref{r_criterion}) is not fulfilled. Note that $V_{\min }$ has a
sharp peak in the critical range of bistability. Anyway in the
far-above threshold region the CV entanglement of the system is
guaranteed. More exactly, the asymptotic value of $V_{\min }$ for
all detunings is equal to $0.75$.

\begin{figure}
\includegraphics[angle=-90,width=0.48\textwidth]{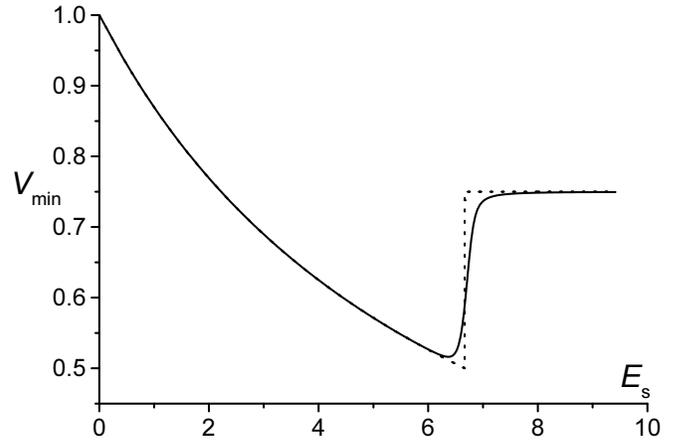}
\caption{Illustration of the critical region squeezing for small values of $%
\protect\kappa /\protect\gamma $ ratio. $\kappa /\gamma =0.5$ for
the solid line, and $\kappa /\gamma=10^{-6}$ for the dot line.}
\label{k_gamma_illustration_fig}
\end{figure}

Attentive reader of this article must be somewhat confused by the
unnatural value of $\kappa /\gamma $ ratio selected to plot the
figures. We emphasize that the choice of parameters is conditioned
merely by illustrative
purposes. For smaller, experimentally available values of the parameter $%
\kappa /\gamma \thicksim 10^{-6}\div 10^{-8}$, the behavior of
$V_{\min }$ is not changed qualitatively in comparison with the
results of Fig.\ \ref {variance_fig}. To illustrate this fact we
plot in Fig.\ \ref {k_gamma_illustration_fig} the behavior of
$V_{\min }$ in the interjacent regime and for two values of
$\kappa /\gamma $. One can see that for the more realistic
parameter value $\kappa /\gamma =10^{-6}$ behavior in the vicinity
of the critical point is more abrupt, which is just a result of
critical region being much narrower. For the bistable dynamics
this narrowing of the critical region makes the peak of $V_{\min
}$ (Fig.\ \ref {variance_fig}, curve 3) much sharper, the rest of
the plot remaining unchanged.

It is remarkable that equations (\ref{r_expression}),
(\ref{a1a2_altform}) allow to explain qualitatively and even
quantitatively the numerical results obtained.

1. In the far below-threshold regime mean photon number can be
represented linear in $p$: $n=p/2\left| \Lambda +1\right| ^{2}\ll
1$. Using this formula it is easy to check that the condition
$V_{\min }<1$ is fulfilled for any $\Lambda $ within the entire
domain where the weak pump
approximation is valid.

2. In the far above-threshold range ($p\gg \left| \Lambda \right|
^{2}$) we can make use of the semiclassical expression
(\ref{meanphotonclassical}) together with (\ref{a1a2}) to study
the variance $V_{\min }$. To this end we write $n=n_{cl}+\delta
n$, and neglect $\delta n$ only where that doesn't lead to loss of
accuracy in the formula (\ref{r_expression}). After some algebra
(see Appendix \ref{appendixa}) we arrive at the following
expression:
\begin{equation}
\lim_{p\rightarrow \infty } V_{\min }=0.5-2\delta n. \label{Vmin
asymptotic form}
\end{equation}
Straightforward, but complicated analytical calculations on the
formula (\ref
{meanphotonexact}) show that $\delta n\rightarrow -0.125$ in the limit \ $%
p\rightarrow \infty $, which leads to the asymptotic value
$V_{\min }=0.75<1 $ (find  in Appendix \ref{appendixb} several
hints regarding our analysis of $\delta n$). Therefore, as our
analysis shows with allowance for quantum fluctuations of
arbitrary level, CV entanglement is always achieved in NOPO above
threshold. What is remarkable, is that a very small quantum
correction $\delta n$ to the semiclassical intracavity photon
number plays an essential role to the production of entanglement
in high-intensity level. Obtained result seems interesting as
provides the example of preserving CV entanglement in the
high-intensity domain.

3. In the near-threshold and bistability regimes analytical study
is too complicated. We have carried out strict analysis only for
the monostable case ($\Delta \Delta _{3}<\gamma \gamma _{3}$),
assuming $\kappa /\gamma \ll 1$. We perform expansion of $V_{\min
}$ around the minimum point $p_{\min }$ in powers of parameter
$s:=\left| \Lambda \right| ^{-1/2}\,\left( s\sim \kappa /\gamma
\right) $ keeping only the significant terms (technics similar to
those described in Appendix \ref{appendixb} is used). The result,
expressed through the pump field intensity $I\sim \left| E\right|
^{2}$, is the following:
\begin{equation}
V_{\min }=0.5+c^{3}f_{1}(c)s+\frac{f_{2}(c)}{s}\left( \frac{I-I_{\min }}{%
I_{th}}\right) ^{2},  \label{Vmin minimum}
\end{equation}
where
\begin{equation}
I_{\min }=I_{th}\left[ 1+f_{3}\left( c\right) s\right],
\label{Imin_expression}
\end{equation}
$c=\sqrt{\cos (\arg \Lambda )}$, $f_{1}(c)>0.0164$,
$f_{2}(c)=0.113+0.00221c-0.330c^{2}+0.371c^{3}-0.132c^{4}$,
$f_{3}\left( c\right) =-2.219+0.217c+2.83c^{2}$. Note that for
fixed damping rates $\gamma_{i}$ $c$ is a decreasing function of
detuning, and $c=1$ at exact resonance. Formula (\ref{Vmin
minimum}) is valid for $c^{5}\gg s$\ (i.e. not too close to the
interjacent regime $c=0$) and $\left| \left( I-I_{\min }\right)
/I_{th}\right| \ll s^{2}$.

Let us discuss the results (\ref{Vmin minimum}),
(\ref{Imin_expression}) in more detail. Positiveness of $f_{1}(c)$
means that the minimal value of $V_{\min }$ doesn't drop below
$0.5 $. This result differs from the studied case of the perfect
two-mode squeezing generated in an undamped NOPA, where $V_{\min
}$ vanishes. Evidently, in our system the larger limiting value
for $V_{\min }$ is due to dissipation and cavity feedback effects.
Then we see that the point of maximal two-mode squeezing $I_{\min
}$ is located close to the generation threshold $I_{th}$. Function
$f_{3}(c)$ is negative for values of $c$ close to zero, however it
increases and becomes positive as $c$ approaches 1. Thus the point
of maximal two-mode squeezing can be located both below and above
the generation threshold of monostable dynamics.

Formulae (\ref{Vmin minimum}) and (\ref{Imin_expression}) are an
example of a result that couldn't be obtained with the help of
linear treatment of quantum fluctuations. Although the linearized
theory gives correct expressions for $V_{min}$ in the below- and
above-threshold ranges, it not only fails to link those solutions
through the narrow range of the generation threshold of monostable
regime, but also turns entirely invalid in the interjacent and
bistable regimes.

In the light of experimental investigations of CV quadrature
entanglement in NOPA and NOPO \cite{Zhang, Zhang_OpticsExpress},
the obtained value for the highest achievable squeezing of 3\,dB
($V_{\min }=0.5$) may seem inadequate. Indeed, the measurements of
the frequency spectra of the variances of the output
quadrature-phase amplitudes reveal up to 3.7\,dB squeezing for
NOPA \cite{Zhang}, and up to 4.9\,dB squeezing for NOPO operating
above threshold \cite{Zhang_OpticsExpress}. However, we remind
that our results pertain to the full squeezing of the system and
not the spectral component squeezing. Moreover, as noted in
\cite{Mandel_JOSA}, states with even perfect squeezing of a
certain spectral component may be non-squeezed in the full sense.
Note, that complete study of that question is beyond the scope of
this paper.

\begin{figure}
\includegraphics[angle=-90,width=0.48\textwidth]{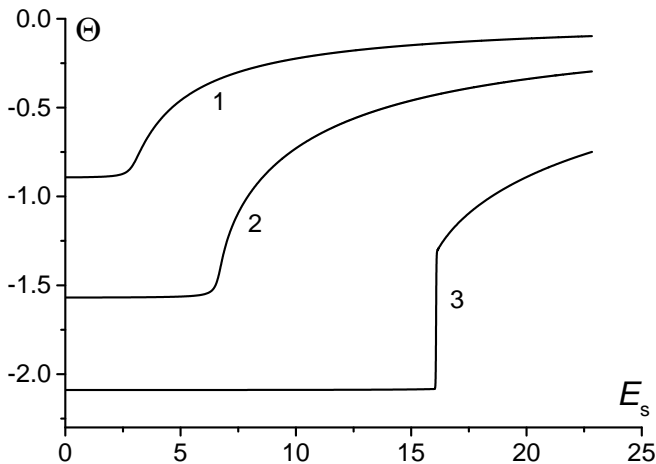}
\caption{$\Theta =\arg \langle a_{1}a_{2}\rangle - \Phi _{E}$
versus dimensionless amplitude of the pump field $E_{s}$ for the
same parameters as for Fig.\ \ref{intensity_fig}.}
\label{theta_fig}
\end{figure}

Our results allow us to explain qualitatively the growth of
$V_{\min }$ in the critical region of bistability. We remind that
the minimization of the variance $V$ is performed through a
specific selection of phases of the quadrature amplitudes $X_{k}$
and $Y_{k}$ - see formula (\ref{minimizing_phases}). In the
bistable regime the state can be regarded as a mixture
$\hat{\rho}=p_{w}\hat{\rho}_{w}+p_{s}\hat{\rho}_{s}$ of two states
corresponding to the weakly and strongly excited semiclassical
stable solutions. What if both $\hat{\rho}_{w}$ and
$\hat{\rho}_{s}$ are highly squeezed two-mode states, but the
minimizing phases are rather different for them? In this case the
minimal variance for their mixture will be achieved for an
intermediate value of the phases, which is good for none of them.
The analysis of the phase of $\langle a_{1}a_{2}\rangle $ suggests
that this is indeed the reason. Fig.\ \ref{theta_fig} shows that
the minimizing phase changes very abruptly in the bistable
critical region. In this context we note that in the critical
bistable regime the state is extremely different from Gaussian and
therefore the condition $V_{\min }>1$ does not witness against
entanglement since the criterion \ref{r_criterion} provides only
sufficient condition for CV entanglement in this case.

In summary, we have investigated CV entangling resources for all
operational regimes of nondegenerate OPO in the most basic and
explicit way through the P-complex probability distribution. We
have studied the entanglement as two-mode squeezing and have shown
that entanglement is present in the system for quantum noise of
arbitrary level and, what is remarkable, for a wide range of
intensity of the coherent driving field including the
far-above-threshold limit. The numerical and analytical analysis
for the quantum critical range of NOPO has provided a clear
distinction between the degree of two-mode squeezing for
monostable and bistable regimes. Most favorable is the monostable
regime where CV\ entanglement can be maintained for all values of
the pump intensity and for experimentally available parameters. In
our study we have not analyzed any entanglement measure which is
extremely difficult to handle analytically for non-Gaussian states
\cite{endnote1}. This topic is currently being explored and will
be the subject of forthcoming work. We hope the results obtained
can serve as a guide for further theoretical and experimental
studies of bright CV entangled light.
\begin{acknowledgments}
Acknowledgements: We gratefully acknowledge useful discussions
with H.\ H.\ Adamyan, H.\ K.\ Avetissyan, J.\ Bergou and C.\ W.\
Gardiner. This work was supported by the NFSAT PH 098-02/CRDF
grant no. 12052 and the ISTC grant no. A-353.
\end{acknowledgments}

\appendix

\section{Asymptotic form of $V_{\min}$}\label{appendixa}

Supposing that for $p\gg \left| \Lambda \right| ^{2}$ $\delta n$
is monotonous and $\delta n \ll n_{cl}$ (these assumptions are
justified during a later analysis) it can be easily shown that
$n\approx \frac{1}{2} \sqrt{p}$ and
\begin{equation}
\lim_{p\rightarrow \infty}\frac{p}{n}\frac{\partial n}{\partial
p}=\frac{1}{2}, \label{pndndp}
\end{equation}

The following exact expression is also not difficult to obtain
using formula (\ref{a1a2_altform}):
\begin{equation}
\left| \left\langle a_{1}a_{2}\right\rangle \right| =n\sqrt{1 +
\frac{Q}{p}\left( 2\delta n+\frac{p}{n}\frac{\partial n}{\partial
p}\right) }, \label{modulea1a2}
\end{equation}
where
\begin{equation}
Q=2\sqrt{p-\left( \mathop{\rm Im}\Lambda \right) ^{2}}+2\delta
n+\frac{p}{n}\frac{\partial n}{\partial p}.
\end{equation}

$Q\approx 2\sqrt{p}$, when $p\rightarrow \infty$. Therefore
(\ref{modulea1a2}) can be expanded into series, whereupon we find
\begin{equation}
\left| \left\langle a_{1}a_{2}\right\rangle
\right|=n+\frac{nQ}{2p}\left( 2\delta n+\frac{p}{n}\frac{\partial
n}{\partial p} \right)+O(p^{-\frac{1}{2}}),
\end{equation}

This immediately yields the asymptotic behavior (\ref{Vmin
asymptotic form}) of $V_{\min }$ .

\section{Asymptotic value of $\delta n$} \label{appendixb}

We decompose the value of $2n_{cl}$ into the integer and fraction
parts: $2n_{cl}=\mu+\xi$, with $\mu =E\left[ 2n_{cl}\right] $ and
$\xi=F\{2n_{cl}\}$. After that we can write:
\begin{equation}
\delta n=n-n_{cl}=\frac{F}{2N}-\frac{\xi}{2},
\end{equation}
where
\begin{equation}
F=\sum_{j=0}^{\infty }\left( j-\mu \right) N_{j},
\end{equation}
and $N$ and $N_{j}$ are defined with formula (\ref{Nexpression}).

It can be shown that $N_{\mu}$  is the largest term in $N$ and
that the values of $N$ and $F$ can be calculated with sufficient
precision by summing only about $O(\sqrt{\mu})$ terms around
$N_{\mu}$. Within that range of indices the term $N_{\mu+j}$ can
be represented as below:
\begin{equation}
N_{\mu+j}\approx \exp \left[ -\frac{j(j+1-2\xi)}{\mu}\right] .
\end{equation}
Summation of such terms can be safely replaced with integration,
with bounds extended to infinity. We finally obtain the following
expression for $N$:
\begin{equation}
N\approx \sqrt{\pi \mu}\exp \left[\frac{(1-2\xi)^2}{4\mu}\right]
N_{\mu}.
\end{equation}

Calculation of $F$ requires more accurate treatment, since it
contains positive and negative parts. We ensure against errors due
to subtraction of two large values rewriting $F$ as follows:
\begin{equation}
F=\sum_{j=1}^{O(\sqrt{\mu})} j\left( N_{\mu+j}-N_{\mu-j} \right) .
\end{equation}

The difference $N_{\mu+j}-N_{\mu-j}$ can be simplified to an
appropriate approximate form, allowing to replace the sum with
integration. Proceeding like with $N$ we arrive at $\delta n
\displaystyle{\mathop{\rightarrow}_{p\rightarrow \infty}}-0.125$.


\begin{thebibliography}{0}

\bibitem{QITwithCV} Quantum Information Theory with Continuous
Variables, S.\ L.\ Braunstein and A.\ K.\ Pati, eds. (Kluwer,
Dordrecht, 2003), and references therein.

\bibitem{CVexperimental} S.\ L.\ Braunstein and H.\ J.\ Kimble, Phys.\ Rev.\ Lett.\ {\bf 80}, 869 (1998);
A.\ Furusawa, J.\ L.\ Sorensen, S.\ L.\ Braunstein, C.\ A.\ Fuchs,
H.\ J.\ Kimble, and E.\ S.\ Polzik, Science {\bf 282}, 706 (1998);
W.\ P.\ Bowen, N.\ Treps, , R.\ Schnabel, and P.\ K.\ Lam, Phys.\
Rev.\ Lett.\  {\bf 89} 253601 (2002).

\bibitem{Ralph} T.\ C.\ Ralph, Phys.\ Rev.\ A {\bf 61}, 010303(R)
(2000); Ch.\ Silberhorn et al, Phys.\ Rev.\ Lett.\ {\bf 88},
167902 (2002); F.\ Grosshans, G.\ Van Assche, J.\ Wenger, R.\
Brown, N.\ J.\ Cerf, P.\ Grangier, Nature, {\bf 421} 238 (2003).

\bibitem{Reid}  M.\ D.\ Reid and P.\ D.\ Drummond, Phys.\ Rev.\ Lett.\ {\bf %
60}, 2731 (1988); M.\ D.\ Reid, Phys.\ Rev.\ A {\bf 40}, 913
(1989); P.\ D.\ Drummond and M.\ D.\ Reid, Phys.\ Rev.\ A {\bf
41}, 3930, (1990).

\bibitem{OuPereiraKimblePeng}  Z.\ Y.\ Ou, S.\ F.\ Pereira, H.\ J.\ Kimble, and K.\ C.\ Peng,
Phys.\ Rev.\ Lett.\ {\bf 68}, 3663 (1992); S.\ F.\ Pereira, Z.\
Y.\ Ou and H.\ J.\ Kimble, Phys.\ Rev.\ A {\bf 62}, 042311 (2002).

\bibitem{Korolkova} Ch.\ Silberhorn, P.\ K.\ Lam, O.\ Wei\ss, F.\ K\"{o}nig,
N.\ Korolkova, and G.\ Leuchs, Phys.\ Rev.\ Lett.\ {\bf 86}, 4267
(2001).

\bibitem{Mert}  S.\ Reynaud, C.\ Fabre, and E.\ Giacobino, J.\ Opt.\ Soc.\ Am.\ {\bf B4},
1520 (1987); J.\ Mertz, T.\ Debuisschert, A.\ Heidman, C.\ Fabre,
and E.\ Giacobino, Opt.\ Lett.\ {\bf 16}, 1234 (1991); J.\ G.\
Rarity, P.\ R.\ Tapster, J.\ A.\ Levenson, J.\ C.\ Garreau, I.\
Abram, J.\ Mertz, T.\ Debuisschert, A.\ Heidman, C.\ Fabre, and
E.\ Giacobino, Appl.\ Phys.\ B {\bf 55}, 250 (1992); K.\ C.\ Peng,
Q.\ Pan, H.\ Wang, Y.\ Zhang, H.\ Su, and C.\ D.\ Xie, Appl.\
Phys.\ {\bf 66}, 755 (1998).

\bibitem{Lauret} J.\ Lauret, T.\ Coudreau, N.\ Treps, A.\
Ma\^{i}tre, and C.\ Fabre, quant-ph/0304111 (2003).

\bibitem{Zhang}  Y.\ Zhang, H.\ Wang, X.\ Li, J.\ Jing, C.\ Xie, and K.\
Peng, Phys.\ Rev.\ A {\bf 62}, 023813 (2000); X.\ Li, Q.\ Pan, J.\
Jing, J.\ Zhang, C.\ Xie, K.\ Peng, Phys.\ Rev.\ Lett.\ {\bf 88},
047904 (2002).

\bibitem{PRA_65_052306} N.\ Korolkova, G.\ Leuchs, R.\ Loudon, T.\
C.\ Ralph, and C.\ Silberhorn, Phys. Rev A {\bf 65}, 052306
(2002).

\bibitem{KK}  G.\ Yu.\ Kryuchkyan and K.\ V.\ Kheruntsyan, Opt.\ Comm.\ {\bf %
127}, 230 (1996); K.\ V.\ Kheruntsyan, D.\ S.\ Kr\"{a}hmer, G.\
Yu.\ Kryuchkyan, K.\ G. Petrosyan, Opt.\ Comm.\ {\bf 157}, 139
(1997).

\bibitem{McN}  K.\ J.\ McNeil and C.\ W.\ Gardiner, Phys.\ Rev.\ A {\bf 28},
1560, (1983).

\bibitem{KP}  G.\ Yu.\ Kryuchkyan, K.\ G.\ Petrosyan and K.\ V.\
Kheruntsyan, JETP Lett.\ {\bf 63}, 526 (1996).

\bibitem{Drummond} M.\ D.\ Reid, and P.\ D.\ Drummond, Phys.\
Rev.\ A {\bf 40}, 4493 (1989).

\bibitem{NOPOwigner} K.\ V.\ Kheruntsyan, K.\ G.\ Petrosyan,
Phys.\ Rev.\ A {\bf 62}, 015801 (2000).

\bibitem{Duan}  L.\ M.\ Duan, G.\ Giedke, J.\ I.\ Cirac, and P.\ Zoller,
Phys.\ Rev.\ Lett.\ {\bf 84}, 2722 (2000); R.\ Simon Phys.\ Rev.\ Lett.\ {\bf %
84}, 2726 (2000); G.\ Giedke, B.\ Kraus, M.\ Lewenstein, and J.\
I.\ Cirac, Phys.\ Rev.\ Lett.\ {\bf 87}, 167904 (2001); R.\ F.\
Werner and M.\ M.\ Wolf, Phys.\ Rev.\ Lett.\ {\bf 86}, 3658
(2001).

\bibitem{Bowen_PRL_90} W.\ P.\ Bowen, R.\ Schnabel, P.\ K.\ Lam, and T.\ C.\
Ralph, Phys.\ Rev.\ Lett.\ {\bf 90}, 043601 (2003).

\bibitem{Zhang_OpticsExpress} Y.\ Zhang, K.\ Kasai, M.\ Watanabe,
Optics Express, {\bf 11}, 14 (2003).

\bibitem{Mandel_JOSA} Z.\ Y.\ Ou, C.\ K.\ Hong, and L. Mandel, J.\
Opt.\ Soc.\ Am.\ B {\bf 4}, 1574 (1987).

\bibitem{endnote1} The elaboration and study of several entanglement measures has been
the subject of recent works. See, for example: G.\ Vidal, R.\ F.\
Werner, Phys.\ Rev.\ A {\bf 65}, 032314 (2002), and references
therein.

\end{thebibliography}
\end{document}